\documentclass[conference]{IEEEtran}
\IEEEoverridecommandlockouts
\usepackage{cite}
\usepackage{amsmath,amssymb,amsfonts}

\usepackage{algorithmic}
\usepackage{textcomp}
\usepackage{xcolor}
\usepackage{graphicx}
\usepackage{url}
\usepackage{subfig}
\usepackage{subcaption}
\usepackage{booktabs}
\usepackage{caption}
\usepackage{multirow}
\usepackage{array}
\usepackage{float}
\usepackage{hyperref}
\usepackage{pifont}
\usepackage{listings}
\usepackage{balance}
\usepackage{cmap}  

\begin{document}

\title{Is LLM-Generated Code More Maintainable \& Reliable than Human-Written Code?}

 \author{
     \IEEEauthorblockN{
         Alfred Santa Molison\IEEEauthorrefmark{1}, 
         Marcia Moraes\IEEEauthorrefmark{2}, 
         Glaucia Melo\IEEEauthorrefmark{1}, 
         Fabio Santos\IEEEauthorrefmark{2}, 
         Wesley K. G. Assunção\IEEEauthorrefmark{3}
     } 
     \IEEEauthorblockA{\IEEEauthorrefmark{1}\textit{Toronto Metropolitan University}, Toronto, Canada \\
     alfredsantamolison@torontomu.ca, glaucia@torontomu.ca}
     \IEEEauthorblockA{\IEEEauthorrefmark{2}\textit{Colorado State University}, Fort Collins, USA \\
     marcia.moraes@colostate.edu, fabio.deabreusantos@colostate.edu}
     \IEEEauthorblockA{\IEEEauthorrefmark{3}\textit{North Carolina State University}, Raleigh, USA \\
     wguezas@ncsu.edu}
 }


\IEEEoverridecommandlockouts
\IEEEpubid{\makebox[\columnwidth]{979-8-3315-9147-2/25/\$31.00 ©2025 IEEE} \hspace{\columnsep}\makebox[\columnwidth]{ }}

\maketitle
\begin{abstract}
Background: The rise of Large Language Models (LLMs) in software development has opened new possibilities for code generation. Despite the widespread use of this technology, it remains unclear how well LLMs generate code solutions in terms of software quality and how they compare to human-written code. Aims: This study compares the internal quality attributes of LLM-generated and human-written code. Method: Our empirical study integrates datasets of coding tasks, three LLM configurations (zero-shot, few-shot, and fine-tuning), and SonarQube to assess software quality. The dataset comprises Python code solutions across three difficulty levels: introductory, interview, and competition. We analyzed key code quality metrics, including maintainability and reliability, and the estimated effort required to resolve code issues. Results: Our analysis shows that LLM-generated code has fewer bugs and requires less effort to fix them overall. Interestingly, fine-tuned models reduced the prevalence of high-severity issues, such as blocker and critical bugs, and shifted them to lower-severity categories, but decreased the model's performance. In competition-level problems, the LLM solutions sometimes introduce structural issues that are not present in human-written code. Conclusion: Our findings provide valuable insights into the quality of LLM-generated code; however, the introduction of critical issues in more complex scenarios highlights the need for a systematic evaluation and validation of LLM solutions. Our work deepens the understanding of the strengths and limitations of LLMs for code generation.
 \end{abstract}


\section{Introduction}


The rapid advancement of Large Language Models (LLMs), such as OpenAI's ChatGPT, has led to their increasing application in various domains~\cite{Bahrini2023}.
These models have demonstrated remarkable capabilities in understanding and generating human-like text~\cite{yao2019plan}. In software development, LLMs have opened up new possibilities, with promising results for code generation~\cite{finnie2022robots,denny2023conversing, savelka2023can}. As this new technology becomes more prevalent in software development workflows~\cite{10.1145/3695988, 10.1145/3631520, 10.1145/3663529.3663839}, assessing its performance in generating high-quality code and comparing it with human-generated code is crucial, particularly in terms of code quality issues~\cite {xu2024distinguishing}. 

Previous studies on the capabilities of LLMs in code generation focused mainly on model performance 
\cite{10.1145/3587102.3588792,10578820,10.1145/3587102.3588794,10.1145/3649405.3659473}, not considering aspects of code quality. Jamil et al. \cite{Jamil2025} have benchmarked LLM code generation against human-written code using static quality metrics (e.g., Pylint, Radon) and multi-criteria ranking methods (e.g., TOPSIS). Our study advances the state of the art by incorporating task difficulty, prompt strategy, and post-generation fix effort through SonarQube, providing an actionable analysis of LLM code quality in realistic development settings. 

Our study centers on the following research question (RQ): \textit{Is LLM-Generated Code More Maintainable and Reliable than Human-Written Code?}
This RQ quantifies maintainability and reliability differences between code generated by a popular LLM and code written by humans. The robustness of our study lies in comparing code using established software quality metrics with a standardized dataset of coding questions (composed of different difficulty levels)~\cite{hendrycksapps2021}. We analyzed code generated by the GPT API using various learning strategies, namely zero-shot and few-shot prompts, and a fine-tuned model. The LLM-generated code is then compared with the code snippets in the APPS dataset~\cite{hendrycksapps2021}, and the quality metrics of all are analyzed.

The results of our study show that LLM-generated code has fewer bugs and often requires less effort to fix them. Interestingly, fine-tuned models reduced the prevalence of high-severity issues, such as blocker and critical bugs, and shifted them to lower-severity categories. In competition-level problems, the LLM solutions sometimes introduce critical structural issues that are not present in human-written code. However, our results show that the fine-tuned model solutions lost performance (0.47) compared to the few-shot model (0.86) in Pass@1, while we observed the highest performance in human-generated solutions.

Our contributions include a comparative analysis of the maintainability and reliability of human-written and LLM-generated code, as well as the effort time and code performance. Additionally, we make an experimentation infrastructure available that can be used in future studies, replications, or to automate code quality analysis (replication package available in the repository~\cite{SupplementaryMaterial}). Furthermore, our findings provide valuable insights into the quality of LLM-generated code. Due to the introduction of critical issues in more complex cases, our study emphasizes the need for systematic evaluation and validation of LLM solutions. Our work deepens the understanding of the strengths and limitations of LLMs in automated code generation by providing a systematic analysis focused on maintainability and reliability issues, done through the use of a novel experimental infrastructure. Furthermore, with the methodology and results of our study, we are one step closer to establishing evidence-based guidelines for creating more efficient quality assessment frameworks for academic and industry environments.


\section{Background and Related Work} \label{background}

The integration of Artificial Intelligence (AI) into software development has significantly advanced in recent years~\cite{sauvola2024future}. This section provides an overview of relevant literature and foundational concepts that support this study.

\subsection{LLMs in Code Generation}
LLMs leverage advanced natural language processing capabilities to facilitate the automatic generation of source code, significantly enhancing developer productivity~\cite{Leung2023}. Tosi~\cite{Tosi2024} highlights that integrating AI generative models in software development optimizes coding practices and redefines the overall software engineering landscape. 
However, while studies like Kwon et al.'s~\cite{Kwon23} research on automated program repair demonstrate the utility of LLMs in debugging and patching buggy scripts, they primarily focus on correctness rather than broader software quality metrics such as maintainability or reliability. Recent empirical evaluations provide further insights into these aspects. For instance, Siddiq et al.~\cite{siddiq2024quality} and Moratis et al.~\cite{moratis2024write} identified code smells and harmful patterns in LLM-generated outputs. However, their findings lack detailed numerical benchmarks to assess long-term maintainability or adherence to software quality standards. Recent work by Abbassi et al.~\cite{abbassi2025unveiling} introduces a taxonomy of inefficiencies in LLM-generated code, highlighting that maintainability issues, such as unnecessary complexity and improper documentation, are prevalent, but vary significantly across models depending on their training data. These findings underscore the critical need for quantitative evaluations of maintainability and reliability in LLM-generated code to ensure its suitability for production environments. As developers increasingly rely on LLMs for code generation~\cite{husein2024large}, understanding these nuances becomes crucial for maintaining software quality and ensuring that generated code meets the required standards.

Leung et al.~\cite{Leung2023} emphasize the importance of developer-LLM interaction interfaces, which facilitate seamless collaboration between human expertise and machine-generated code. This collaboration not only enhances software development efficiency but also raises questions about the quality and security of LLM-generated code. 

\subsection{Software Quality Metrics}

Quality metrics play an important role in assessing and ensuring the quality of software products~\cite{Trautsch2021, Kandengwa2021}. 
Metrics can facilitate stakeholder communication, providing a common language to discuss software quality and project status. They are particularly relevant in Agile environments, where teams use metrics to assess project success and adapt processes~\cite{Molnar2020}. 

Static code analysis tools, such as SonarQube, are essential to assess software quality, as they provide objective, automated code evaluation without the need for execution~\cite{AlOmar2023}. This analysis helps identify potential vulnerabilities, adhere to coding standards, and improve overall code maintainability. 
SonarQube\footnote{\url{http://sonarsource.com}} is a cloud-based platform that provides static code analysis tools to help developers maintain and improve code quality. 
By integrating SonarQube into the development workflow, teams can analyze their codebase, ensuring that quality checks are applied consistently throughout the software development lifecycle~\cite{Dabdawb2021,Shatnawi2020}. 

In this study, we leverage an infrastructure (presented in Section~\ref{metholodogysec}) combining LLM  and SonarQube to gather metrics to automatically evaluate generated code. Organizations can benefit from such infrastructure and foster a quality culture, reduce defects, and deliver more reliable software products to their users.

\subsection{Previous Studies on AI-Generated Code Quality}

The quality of AI-generated code is an emerging area of research. 
Recent studies have empirically evaluated the correctness and quality of the code that various LLMs produce \cite{LiuFinger2024,Coignion2024,Nascimento2023}. These studies' findings revealed significant variations in code quality, 
indicating a need for improved evaluation metrics and methodologies to assess the quality of AI-generated code comprehensively \cite{Tosi2024}. 
Other studies suggest that while LLMs can enhance code generation, they still struggle to produce maintainable and high-quality code~\cite{Liu2024Refining}. 
Golendukhina et al.~\cite{Golendukhina2022} highlight the lack of training and guidelines for AI engineers on code quality and testing. 
The gap in training and established quality standards is crucial as AI integration in software development processes continues to grow. The current literature suggests a pressing need for comprehensive frameworks that evaluate the quality of AI-generated code and provide actionable insights to developers to enhance the reliability and maintainability of such code \cite{khankhoje2023quality,marar2023advancements}. 

To harness the full potential of AI in software engineering, developers must ensure that the generated code meets the necessary quality benchmarks. 
Moratis et al.~\cite{moratis2024write} concluded that 11.9\% of the code created by ChatGPT can harm the application. 
Sajadi et al.~\cite{sajadi2025llmsconsidersecurityempirical} qualitatively analyzed Python and JavaScript code from DevGPT dataset, 
looking specifically into verifying if LLMs can spot security vulnerabilities from code extracted from StackOverflow. 
Dantas et al.~\cite{dantas2023} focused on comparing ChatGPT-generated Java code snippets with 365 CROKAGE recommendations from StackOverflow, finding that ChatGPT produced cleaner code with more consistent naming conventions. 
Elgedawy et al.~\cite{elgedawy2024ocassionally} compared code generated by LLMs to address nine types of tasks to assess security issues, including whether expressing security consciousness is present or not in the prompt. 
Yetiştiren et al. \cite{yetiştiren2023evaluatingcodequalityaiassisted} explored 164 questions from HumanEval dataset \cite{chen2021evaluating}, investigating reliability, maintainability, security, validity, and correctness using the default setup of code assistants: GitHub Copilot, Amazon CodeWhisperer, and ChatGPT. The study revealed that ChatGPT achieved the highest correctness and validity. Amazon CodeWhisperer showed the lowest technical debt and also the lowest correctness. Github Copilot demonstrated high validity but had more reliability issues in comparison to the other tools.

Altogether, these studies inform our choices of metrics to analyze and compare LLMs and human-generated code, which are reliability, maintainability, and security using zero-shot, few-shot, and fine-tuning methods. 
As an extension to previous works, our pipeline is based on SonarQube, uses diverse metrics, and a large dataset (comprising interview and competition questions in Python) to compare with human-generated code.

\section{Methodology} \label{metholodogysec}

To systematically analyze and compare the quality of LLM-generated and human-written code, we built an experimentation infrastructure, prepared the dataset, and used analytical tools to derive measurable software quality outcomes. Scripts used for preprocessing and analysis are included in the replication package repository~\cite{SupplementaryMaterial}.

\begin{figure}[!tp]
    \centering
    \includegraphics[width=1\columnwidth]{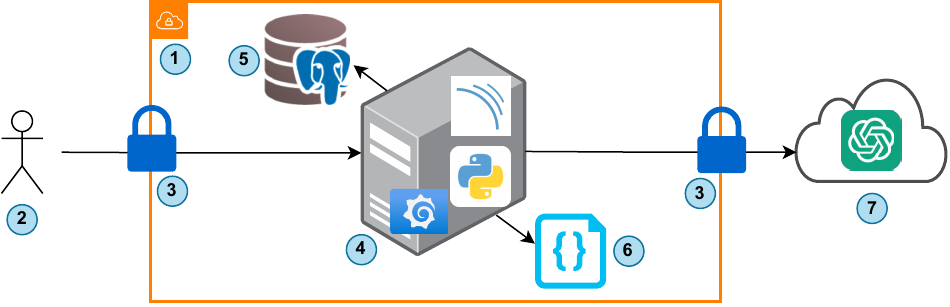}
    \caption{Overview of the experimentation architecture.}
    \label{fig:architecture}
\end{figure}

The experimentation infrastructure consists of a Linux virtual machine (VM) running Ubuntu, Python, Postgres, and SonarQube. This infrastructure supports our methodology, as it systematically integrates machine learning, database management, and software quality analysis within a secure and controlled environment. 
Figure \ref{fig:architecture} overviews the high-level architecture for our infrastructure. 
The components are labelled with numbers corresponding to specific functionalities and elements in the setup, described below.

\textit{\ding{202}~VPC (Virtual Private Cloud):} The experimentation infrastructure is enclosed within a secure and isolated virtual private cloud environment, ensuring controlled access and protected communication among all components. 
\textit{\ding{203}~Users:} The users of this infrastructure, represented as individuals or entities, execute the experiments by securely accessing resources within the VPC. 
\textit{\ding{204}~Restricted Access:} Access to the system is secured using authentication mechanisms, including username-password credentials and API tokens. These ensure authorized personnel or services can connect to the system or its components. 
\textit{\ding{205}~Server Instance:} A central server instance is depicted, which runs multiple key services, including SonarQube for software quality analysis, Grafana for visualization, and Python for scripting and data processing. 
\textit{\ding{206}~PostgreSQL Database:} The database server has a dual purpose: it manages SonarQube data and stores the datasets and the results of the analyses. 
\textit{\ding{207}~Python Source Code Files:} Solution code files are generated during the analysis pipeline (i.e., LLM-generated code). These files are stored and prepared for evaluation by SonarQube. 
\textit{\ding{208}~ChatGPT API:} A connection to the ChatGPT API allows the system to send coding questions and receive LLM-generated solutions. These solutions are subsequently processed and analyzed in parallel with human-written code.

\subsection{Experimentation workflow}\label{Experimentationdatasetmethod}

\begin{figure}[!tp]
    \centering
    \includegraphics[width=1.05\columnwidth]{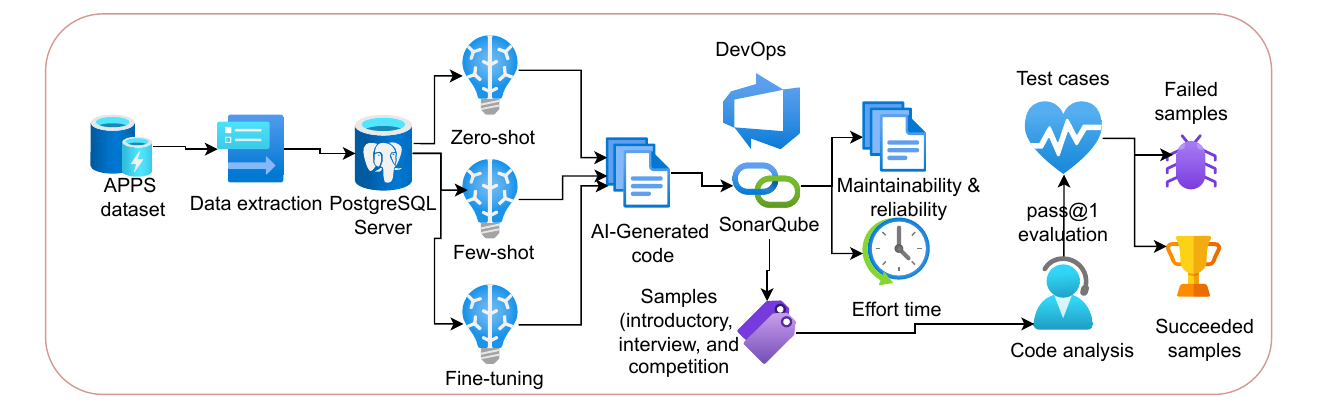}
    \caption{Overview of the experimentation workflow.}
    \label{fig:method}
\end{figure}

Figure~\ref{fig:method} presents an overview of the experimentation workflow of our empirical study. The input data for our work is a human-written dataset~\cite{hendrycksapps2021}, which was curated and stored into PostgreSQL (details in Section~\ref{datasetmethod}). The question descriptions from this dataset were used to obtain the LLM-generate solutions (details in Section~\ref{llmselec}).
Then, SonarQube was used to analyze four datasets (presented in Section~\ref{sonar}) containing code snippets from the original dataset and LLM-generated datasets. 
The results from SonarQube were used for the quantitative comparison of maintainability and reliability between the human-written and LLM-generated datasets (described in Section~\ref{comparativeana}). 
Lastly, we sample some LLM-generated solutions to perform an in-depth qualitative analysis, with the goal of gaining insights into the strengths and weaknesses of the LLM during code generation.

\subsection{Dataset Description} \label{datasetmethod}
\begin{figure}[ht!]
    \centering
    \includegraphics[width=1\linewidth]{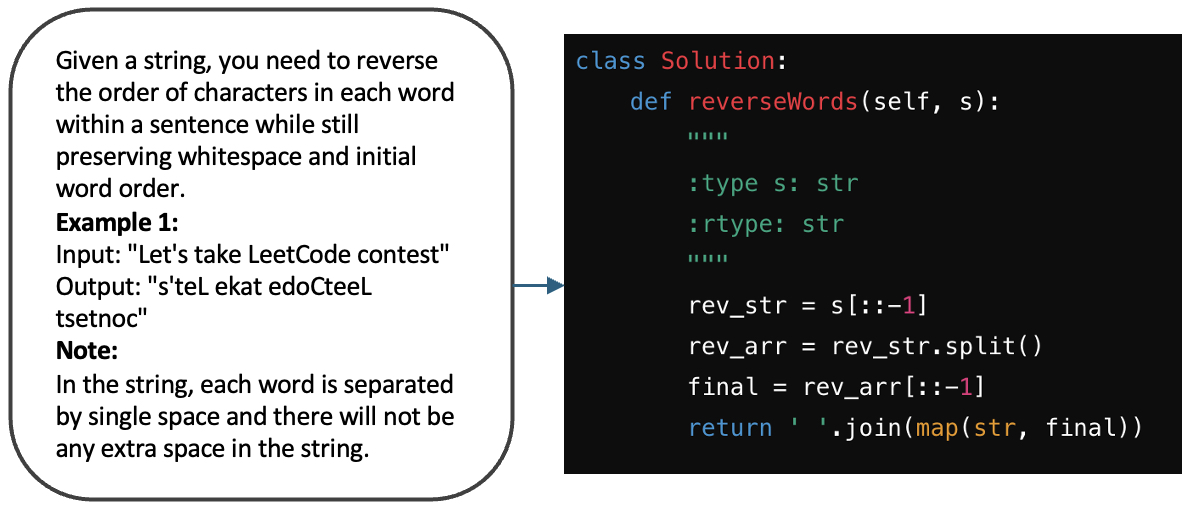}
    \caption{Question (Problem) and Answer (Solution).}
    \label{fig:questionAnswer}
\end{figure}

The dataset used in our study~\cite{hendrycksapps2021} contains 10,000 coding questions (problem descriptions) and solutions (code snippets) written in Python. The questions are categorized as introductory, interview, and competition difficulty levels. This same dataset has also been used in various studies~\cite{jiang2024survey,sharma2023stochastic} and includes multiple solutions for each coding question. The questions are across various programming topics, making it a valuable resource for analyzing problem-solving patterns and evaluating code quality. Figure~\ref{fig:questionAnswer} presents an example of a question and a possible solution from the dataset.

The coding solutions are written by humans, collected from coding websites like Codeforces, AtCoder, Kattis, and Codewars;\footnote{Codeforces: \url{https://codeforces.com/}, AtCoder: \url{https://atcoder.jp/}, Kattis: \url{https://open.kattis.com/}, and Codewars: \url{https://www.codewars.com/}} which are public platforms where users post and solve programming problems. 
The problems were curated during six months by a group of graduate and undergraduate students to remove questions that rely on images and duplicates, standardize format and test cases, set a category, and test them using a designed test framework. 
The difficulty levels are: \textit{introductory-level} problems with straightforward tasks solvable by programmers with 1–2 years of experience, such as counting vowels or calculating running sums. \textit{Interview-level} problems are more algorithmic and, commonly seen in technical interviews, involve data structures like trees and graphs. Lastly, \textit{competition-level} problems are the most challenging and designed for advanced programming competitions \cite{hendrycksapps2021}. 

This dataset is ideal for conducting studies with LLMs. The problem descriptions allow us to experiment with zero-shot learning, without the need for preexisting code snippets. The availability of code solutions facilitates both few-shot learning and the training of a fine-tuned model.
Furthermore, the dataset's format and diversity of questions and solutions make it an excellent benchmark for studying the code quality generated by foundation models. 



\subsection{LLM Configuration and Code Generation Process} \label{llmselec}

The LLM used in this study is GPT-4o-2024-08-06.\footnote{\url{https://platform.openai.com/docs/models/gpt-4o}} We looked for the latest model that allowed fine-tuning. The response format text is set to include structured outputs and function calls, ensuring compatibility with our data pipeline. The model configuration included a temperature value equal to 1.00, the default setting, which provides an optimal balance between deterministic and stochastic outputs~\cite{openai_audio_translation}. 
The maximum token limit was kept at its default, 2048 tokens, providing a sufficient context window to handle complex programming problems while ensuring efficient processing and resource utilization.
Top-P sampling was configured to 1.00, ensuring the most probable token distributions were used~\cite{openaidoc01}. 
Additionally, frequency and presence penalties were set to 0.00, preventing any bias against repetitive or new tokens~\cite{openaidoc01}.
These configurations ensured that the model provided well-rounded and contextually relevant solutions.

We employed three prompt engineering techniques to generate the code answers~\cite{rahman2018unified,kadam2020review}: zero-shot, few-shot, and fine-tuning, described in what follows. 

\textbf{Zero-shot} refers to a scenario where the model performs a task without being explicitly trained on examples. The model relies on its general language understanding to infer the correct output. The prompt for zero-shot was: ``\textit{Generate a Python code (only one solution) to solve this question:}'', appended to each coding question (problem) from the dataset. 

\textbf{Few-shot} involves providing the model with examples in the input prompt to demonstrate the task. These examples help the model better understand the task requirements, improving performance compared to zero-shot. 
For the few-shot prompt, we leveraged the human dataset, which contained an array of solutions for each question, to build the prompt. Using this array of solutions, we iteratively appended sample solutions to the prompt to provide context for the AI model. For each question, the process involved adding a ``Question'' section followed by multiple ``Sample Solution \{index\}'' entries, where each solution from the array was included. This process looped through all items in the array to create a comprehensive and detailed prompt. The final prompt string, referred to as fullprompt, consisted of the base prompt, the question, and the appended sample solutions (fullprompt = prompt + question + samplesolution). This structured prompt enhanced the model's ability to nudge the model and guide the model toward task completion~\cite{Cogo2024AIwareTutorialPrompt}. 

\textbf{Fine-tuning} consists of retraining the model on a specific dataset to adapt it for a particular task and optimize performance for the given task. 
For the fine-tuned model, we trained the GPT base model on 80 random questions and answers from each difficulty level from the original dataset, totalling 240 questions and answers for training. Then, we tested the model with 160 questions of each difficulty level, totalling 480 questions for testing. The training loss was 0.21, and the process involved 496,176 tokens, representing the total number of individual units (words, subwords, and characters) processed during the fine-tuning phase. 

We submitted each coding question to the ChatGPT server, obtaining LLM-generated code snippets with the three prompt engineering techniques. Thus, after this process, we have four distinct datasets: (i) the \textit{human dataset}, which serves as the baseline, containing human-generated answers to the initial coding questions, and the LLM-generated datasets, namely (ii) \textit{zero-shot dataset}, (iii) \textit{few-shot dataset}, and (iv) \textit{fine-tuned dataset}. 
For each dataset, we kept track of the problem difficulty (introductory, interview, and competition). These datasets were then submitted to SonarQube for analysis. 

\subsection{SonarQube Analysis Setup} \label{sonar}

SonarQube Free Server edition includes a default quality profile for each supported programming language. This profile implements a standardized set of rules applicable to most projects. Our study used the default quality profile and quality gate settings to analyze Python code.\footnote{\url{https://docs.sonarsource.com/sonarqube-server/10.8/analyzing-source-code/languages/python/}}
We configured the analysis with the default existing rules on the SonarQube client to assess the quality of Python code, which is consistent with prior studies \cite{Gigante2023,Mortara2022}. 

\textbf{The scanning process.} 
Our infrastructure creates a new project through Sonar APIs before each scan, ensuring all files are treated as new code and undergo comprehensive analysis. We developed Python scripts that retrieve source code from PostgreSQL tables and organize them into \texttt{.py} files within specific directories, each corresponding to a dataset and difficulty level. 
Our Python scripts interface with SonarQube using the SonarQube Python library, which leverages the SonarScanner CLI to create new projects and initiate scans. 
The scanning process generates JSON files containing the results. 
The scan results are categorized and displayed on SonarQube's Web UI dashboard. For our statistical analysis, we exported the results from the twelve JSON files, considering the three difficulty levels and four datasets. 

\textbf{SonarQube Metrics.}
We analyzed the software quality metrics from SonarQube to evaluate the effectiveness of both human-written and LLM-generated code. SonarQube detects issues in the code that violate predefined coding rules linked to project quality profiles. Each issue corresponds to a deficiency in clean code attributes, such as maintainability, security, or reliability, essential for ensuring high-quality software. 
The problem types are: (i) \textit{Bugs}, coding errors causing runtime failures or unexpected behavior; (ii) \textit{Vulnerabilities}, potential security risks; and (iii) \textit{Code Smells}, maintainability challenges that make the code harder to understand and maintain. 
SonarQube also computes the \textit{effort-in-minute} metric, which refers to the estimated time needed to fix the code, taking into account the severity, complexity, and number of issues. Effort helps quantify the work required to improve code quality.

SonarQube also classifies issues into four severity levels, based on the issues impact on the application behavior, security, or developer productivity. 
\textit{Blocker} represents the most severe issues, such as critical bugs that can significantly impact the application's behavior in production. Examples include memory leaks or unclosed connections. These require immediate attention to avoid application failures. 
\textit{Critical} severity includes bugs with the probability of affecting production behavior or significant security vulnerabilities like SQL injection. These issues also demand urgent review and resolution to ensure application stability and security. 
\textit{Major} severity issues capture quality flaws that negatively affect developer productivity, such as duplicate code blocks or uncovered code. These issues might not cause immediate failures but can hinder efficient development and long-term maintainability. Lastly, 
\textit{Minor} severity issues represent less significant quality flaws, such as stylistic issues or slightly inefficient code structures. These might not have a profound impact, but can still contribute to code clutter or decreased readability. 

\subsection{Comparative Analysis Approach} \label{comparativeana}

To compare the effort-in-minute metrics between the four datasets, we employed the \textit{Mann-Whitney U test}, followed by \textit{Cliff's delta effect size} test. Cliff's delta magnitude was assessed using the thresholds provided by Romano et al.~\cite{Romano:2006}, i.e., $|d|<0.147$ as ``negligible,'' $0.147 \leq |d|<0.33$ as ``small,'' $0.33 \leq |d|<0.474$ as ``medium,'' and otherwise as ``large.'' We also used the \textit{Odds Ratio} to compare bugs and code smells, using the following equation~\cite{szumilas2010explaining}: $Odds Ratio (OR) = \frac{(a/c)}{(b/d)}$.
Probabilities greater than 1 mean that the first subgroup is more likely to report a type of label, whereas probabilities less than 1 mean that the second group has greater chances~\cite {odds-ratio}. 

To further our analysis of the LLM- and human-generated code, we analyze code samples to determine their performance. Three authors evaluated a sample of the LLM-generated solutions using Pass@1 accuracy. We computed the Pass@1 metric \cite{chen2021evaluating} by $\mathrm{Pass@}k = \mathbb{E}\Bigl[1 - \tfrac{\binom{n - c}{k}}{\binom{n}{k}}\Bigr]$, where $k$~=~1, $n$ is the number of coding solutions, and $c$ is the number of passed solutions out of $n$. 

\section{Results} \label{results}

\begin{figure}[!tp]
\includegraphics[width=1\linewidth]{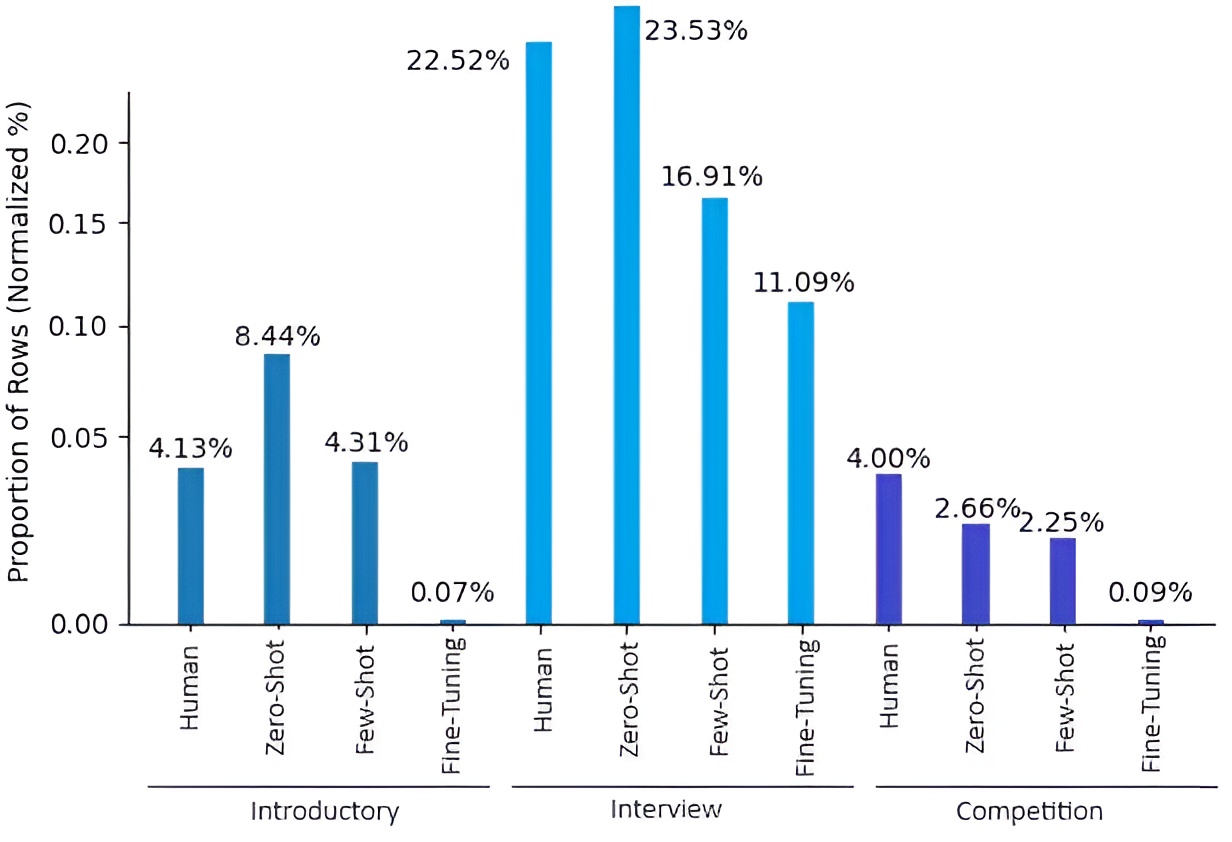}
    \caption{Normalized issue \% per dataset. }
    \label{fig:count_issue}
\end{figure}

Figure \ref{fig:count_issue} presents the normalized percentage of issues identified by SonarQube across code samples generated by humans and the LLM prompt strategies (zero-shot, few-shot, and fine-tuning), classified into task difficulty levels: introductory, interview, and competition. Notably, SonarQube detected no security vulnerabilities in any of the datasets. The chart highlights a stark contrast in issue prevalence depending on the prompting strategy used and task difficulty. Fine-tuning models produced the cleanest code, with nearly negligible issue rates in introductory and competition difficulty levels (0.07\% and 0.09\%, respectively). Zero-shot outputs had the highest issue density among LLM-generated code, especially in introductory and interview tasks. Human-written code produced a higher proportion of issues in the competition difficulty level, and yielded high percentages among the other strategies in the other two difficulty categories.  This analysis compares the total number of rows, normalized as proportions, generated through different approaches. By visualizing the proportion of rows, this analysis highlights the relative data output and issue density from SonarQube for each approach and difficulty level, enabling a better understanding of how different generation techniques impact the distribution of issues within the datasets. This analysis is interesting, as it illustrates the scale and consistency of results across datasets, aiding in the evaluation of the efficacy of AI-driven approaches in generating code compared to human-written solutions.

\subsection{Issue Severity}

\begin{figure*}[!tp]
\includegraphics[width=1\linewidth]{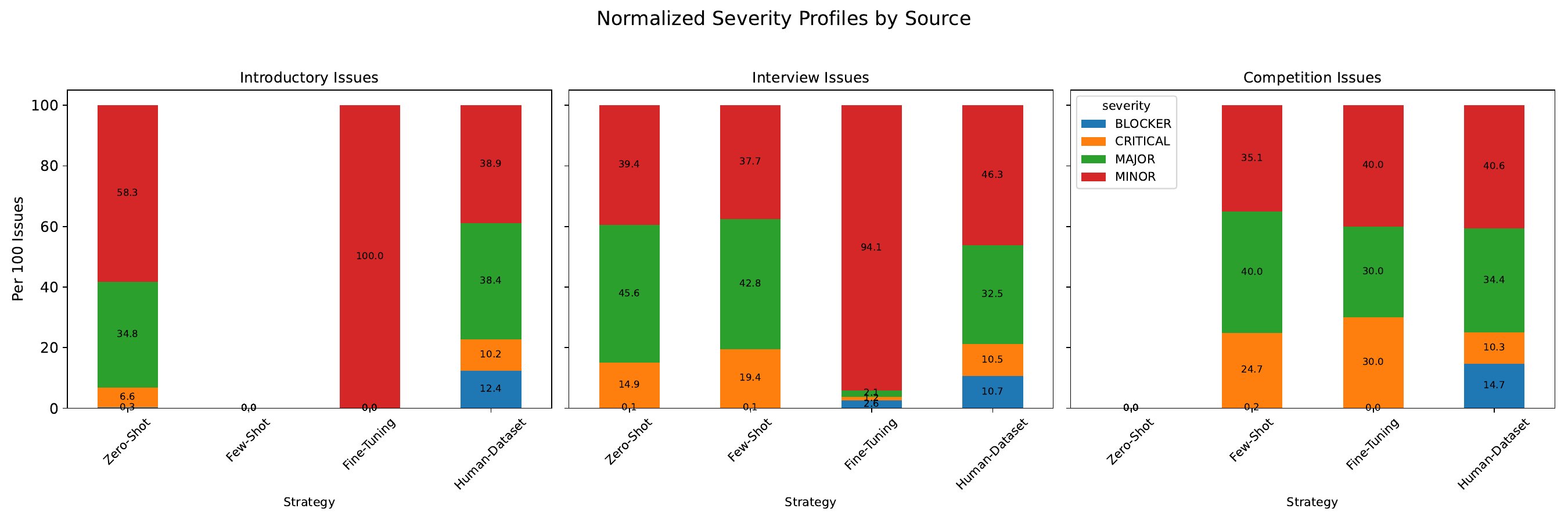}
    \caption{Normalized issues per dataset. }
    \label{fig:severity_distribution}
\end{figure*}


Figure \ref{fig:severity_distribution} reports the severity of issues in each dataset. 
For introductory questions, the human dataset has the highest blocker severity at 12.4\%, indicating significant issues in this baseline.
The other approaches (zero-shot, few-shot, and fine-tuning) nearly eliminate blocker issues, with proportions of 0.3\%, 0.2\% and no occurrence, respectively.

Zero-shot reduces critical severity from 10.2\% to 5.0\%, while few-shot reduces it to 9.8\%. Fine-tuning achieves the lowest critical severity, essentially eliminating it. 
Few-shot shows the highest proportion of major issues at 40.2\%, followed closely by the human dataset at 38.4\%. 
Fine-tuning shows no major issues, reflecting substantial improvement. This learning strategy also has the highest proportion of minor severity at 100\%, outperforming the other learning strategies by reducing or eliminating high-severity issues (blocker, critical, and major) and converting them into minor issues, highlighting its effectiveness in improving system quality. 
Zero-shot performs better than the human dataset, especially in reducing blocker and critical severity, but it is less effective than few-shot and fine-tuning. 

\vspace{1mm}\noindent
\textit{\textbf{Finding 1:} Fine-tuning appears as the most effective approach for introductory questions in improving quality and reducing severity levels to minor. 
Few-shot is a strong intermediate step, showing improvements over both zero-shot and the human dataset, but leaving room for refinement.}
\vspace{1mm}

For the interview questions dataset, zero and few-shot have the lowest proportion of blocker issues (0.1\% for both), while the human dataset approach has the highest (10.7\%). Fine-tuning (2.6\%), zero, and few-shot exhibit improvements over the human dataset but still retain a few blocker issues. 
Fine-tuning exhibits the fewest critical issues (1.2\%), showing a marked improvement over the few-shot (19.4\%), zero-shot (14.9\%), and the human dataset (10.5\%). 
Major issues are significantly reduced in fine-tuning (2.1\%) compared to zero-shot (45.6\%), the human dataset (32.5\%), and few-shot (42.8\%). 
Minor issues dominate in fine-tuning (94.1\%), demonstrating that while most severe issues are addressed, most remaining issues are of low severity. 

\vspace{1mm}\noindent
\textit{\textbf{Finding 2:} The results indicate that fine-tuning consistently reduces the severity of software issues for interview questions, shifting from more severe categories (blocker, critical, and major) to predominantly minor issues. This demonstrates high fine-tuning efficacy in improving software quality compared to other approaches for interview questions.}
\vspace{1mm}

In the competition questions analysis, the human dataset exhibited the most serious issues, with 14.7\% classified as blocker issues. The zero-shot approach significantly reduced this to 0.3\%, while few-shot and fine-tuning approaches eliminated blocker issues. However, the human dataset performed the best for critical issues, with the lowest proportion of critical issues compared to the LLM-generated solutions. In sequence, zero-shot, few-shot, and fine-tuning generated progressively more critical issues, highlighting a tendency for LLM-generated code to introduce more critical-level problems for competition questions than human-generated code. For major issues, the results show that zero-shot and few-shot approaches were tied at 40\%, representing an increase from the human dataset by 34.4\%. Fine-tuning demonstrated a notable improvement by reducing major issues to 30\%.
For minor issues, all LLM solutions reached a reduction compared to the human dataset, which started at 40.6\%. Zero-shot reduced this to 36.6\%, few-shot to 33.5\%, and fine-tuning to 30\%. 

\vspace{1mm}\noindent
\textit{\textbf{Finding 3:} The analysis suggests that while AI approaches, particularly fine-tuning, effectively reduced blocker and major issues, they may introduce more critical issues compared to human-generated code.} 
\vspace{1mm}

To verify the statistical differences, we analyzed the data, aggregating answers (solutions) according to their quality metric (code smell and bug) and severity (blocker, critical, major and minor), resulting in the subgroups presented in Table 
\ref{tab:severities}. 
The human dataset consistently shows the highest percentages of maintainability issues, particularly at the higher difficulty levels. Zero-shot and few-shot methods exhibit moderate maintainability issues, with the zero-shot approach being slightly more prone to such problems.
Fine-tuning emerges as the superior approach, with significantly lower percentages of maintainability issues across all difficulty levels, reaching near-zero levels for introductory and competition questions. For reliability (bugs), the human dataset approach exhibits the most concerning reliability issues, with bug rates ranging from 71.27\% to 87.50\% across the difficulty levels.
Zero-shot and few-shot methods show substantial improvements in reliability compared to the human dataset, with bug rates generally below 5\% for introductory and interview questions. Fine-tuning demonstrates exceptional reliability, achieving 0\% bug rates in multiple instances, particularly for introductory and competition questions. 

\vspace{1mm}\noindent
\textit{\textbf{Finding 4:} Overall, the results suggest that advanced prompt engineering techniques and fine-tuning can outperform the human dataset approach in terms of reducing maintainability and reliability metrics.} 

\begin{table}[!tp]
\caption{Distribution of Severities Per Quality Metric. }
\label{tab:severities}
\resizebox{\columnwidth}{!}{%
\tiny
\begin{tabular}{cc|l|r|r|}
\toprule
\multicolumn{2}{|c|}{\textbf{Dataset}} &
  \multicolumn{1}{c|}{\textbf{\begin{tabular}[c]{@{}c@{}}Severity \\ Type\end{tabular}}} &
  \multicolumn{1}{c|}{\textbf{\begin{tabular}[c]{@{}c@{}}Maintainability\\ Code Smell(\%)\end{tabular}}} &
  \multicolumn{1}{c|}{\textbf{\begin{tabular}[c]{@{}c@{}}Reliability\\Bugs(\%)\end{tabular}}} \\ \midrule
\multicolumn{1}{|c|}{\multirow{14}{*}{Introductory}} & \multirow{4}{*}{\rotatebox[origin=c]{90}{Human}}  & BLOCKER  & 0  & 87.50 \\ 
\multicolumn{1}{|c|}{}                               &                              & CRITICAL & 2.58  & 0  \\ 
\multicolumn{1}{|c|}{}                               &                              & MAJOR    & 9.64 & 1.56  \\ 
\multicolumn{1}{|c|}{}                               &                              & MINOR    & 9.81 & 0  \\ \cmidrule(l){2-5} 
\multicolumn{1}{|c|}{}                               & \multirow{4}{*}{\rotatebox[origin=c]{90}{Zero-Shot}}    & BLOCKER  & 0.06  & 3.13  \\ 
\multicolumn{1}{|c|}{}                               &                              & CRITICAL & 2.58  & 0  \\ 
\multicolumn{1}{|c|}{}                               &                              & MAJOR    & 16 & 4.69  \\ 
\multicolumn{1}{|c|}{}                               &                              & MINOR    & 32 & 0  \\ \cmidrule(l){2-5} 
\multicolumn{1}{|c|}{}                               & \multirow{4}{*}{\rotatebox[origin=c]{90}{Few-Shot}}    & BLOCKER  & 0  & 1.56  \\ 
\multicolumn{1}{|c|}{}                               &                              & CRITICAL & 3  & 0  \\ 
\multicolumn{1}{|c|}{}                               &                              & MAJOR    & 11 & 1.56  \\ 
\multicolumn{1}{|c|}{}                               &                              & MINOR    & 13 & 0  \\ \cmidrule(l){2-5} 
\multicolumn{1}{|c|}{}                               & \multirow{2}{*}{\rotatebox[origin=c]{90}{Fine-T}} & MINOR    & 0.45  & 0  \\ 
\multicolumn{1}{|c|}{}                               &                              & All Others   & 0  & 0  \\ \midrule
\multicolumn{1}{|c|}{\multirow{16}{*}{Interview}}    & \multirow{4}{*}{\rotatebox[origin=c]{90}{Human}}  & BLOCKER  & 0.05  & 71.27 \\ 
\multicolumn{1}{|c|}{}                               &                              & CRITICAL & 3.35  & 0  \\ 
\multicolumn{1}{|c|}{}                               &                              & MAJOR    & 9.82 & 11.33 \\ 
\multicolumn{1}{|c|}{}                               &                              & MINOR    & 14.71 & 0.55  \\ \cmidrule(l){2-5} 
\multicolumn{1}{|c|}{}                               & \multirow{4}{*}{\rotatebox[origin=c]{90}{Zero-Shot}}    & BLOCKER  & 0.03  & 0.28  \\ 
\multicolumn{1}{|c|}{}                               &                              & CRITICAL & 4.94  & 0  \\ 
\multicolumn{1}{|c|}{}                               &                              & MAJOR    & 15.04 & 3.04  \\ 
\multicolumn{1}{|c|}{}                               &                              & MINOR    & 13.08 & 0.55  \\ \cmidrule(l){2-5} 
\multicolumn{1}{|c|}{}                               & \multirow{4}{*}{\rotatebox[origin=c]{90}{Few-Shot}}    & BLOCKER  & 0  & 0.55  \\ 
\multicolumn{1}{|c|}{}                               &                              & CRITICAL & 4.64  & 0  \\ 
\multicolumn{1}{|c|}{}                               &                              & MAJOR    & 10.15 & 1.93  \\ 
\multicolumn{1}{|c|}{}                               &                              & MINOR    & 8.94  & 1.38  \\ \cmidrule(l){2-5} 
\multicolumn{1}{|c|}{}                               & \multirow{4}{*}{\rotatebox[origin=c]{90}{Fine-T}} & BLOCKER  & 0.01  & 8.56  \\ 
\multicolumn{1}{|c|}{}                               &                              & CRITICAL & 0.18  & 0  \\ 
\multicolumn{1}{|c|}{}                               &                              & MAJOR    & 0.30  & 0.55  \\ 
\multicolumn{1}{|c|}{}                               &                              & MINOR    & 14.76 & 0  \\ \midrule
\multicolumn{1}{|c|}{\multirow{14}{*}{Competition}}  & \multirow{4}{*}{\rotatebox[origin=c]{90}{Human}}  & BLOCKER  & 0.11  & 86.30 \\ 
\multicolumn{1}{|c|}{}                               &                              & CRITICAL & 4.96  & 0  \\ 
\multicolumn{1}{|c|}{}                               &                              & MAJOR    & 15.86 & 8.22  \\ 
\multicolumn{1}{|c|}{}                               &                              & MINOR    & 19.49 & 0  \\ \cmidrule(l){2-5} 
\multicolumn{1}{|c|}{}                               & \multirow{4}{*}{\rotatebox[origin=c]{90}{Zero-Shot}}    & BLOCKER  & 0.11  & 0  \\ 
\multicolumn{1}{|c|}{}                               &                              & CRITICAL & 7.38  & 0  \\ 
\multicolumn{1}{|c|}{}                               &                              & MAJOR    & 12.67 & 1.37  \\ 
\multicolumn{1}{|c|}{}                               &                              & MINOR    & 11.67 & 0  \\ \cmidrule(l){2-5} 
\multicolumn{1}{|c|}{}                               & \multirow{3}{*}{\rotatebox[origin=c]{90}{Few-Shot}}    & CRITICAL & 7.16  & 0  \\ 
\multicolumn{1}{|c|}{}                               &                              & MAJOR    & 10.68 & 1.37  \\ 
\multicolumn{1}{|c|}{}                               &                              & MINOR    & 8.92  & 1.37  \\ \cmidrule(l){2-5} 
\multicolumn{1}{|c|}{}                               & \multirow{3}{*}{\rotatebox[origin=c]{90}{Fine-T}} & CRITICAL & 0.33  & 0  \\ 
\multicolumn{1}{|c|}{}                               &                              & MAJOR    & 0.22  & 1.37  \\ 
\multicolumn{1}{|c|}{}                               &                              & MINOR    & 0.44  & 0  \\ \bottomrule
\end{tabular}%
}
\end{table}
\vspace{1mm}

For the quality metrics (bugs and code smells), we computed the odds ratio to compare the question types between groups: human vs. zero-shot, human vs. few-shot, human vs. fine-tuned, zero-shot vs. few-shot, and zero-shot vs. fine-tuned. 
The odds ratio for the quality metric determines how likely it is to get similar responses from both groups. We used a 2x2 contingency table for each comparison.
As shown in Table~\ref{tab:stat-odds}, we found no statistical difference for bugs when comparing zero-shot with few-shot and zero-shot with fine-tuned (except for introductory questions). Conversely, in the same table, we can observe that the odds of a bug occurring in a zero-shot prompt approach are 0.05\% of the odds of a bug occurring in a human solution for a competition question, 0.8\% for introductory, and 0.6\% for interview questions. 

When comparing code smells, presented in Table~\ref{tab:stat-smells}, the odds ratio also revealed statistical differences. We found that the odds of a code smell occurring in a zero-shot prompt approach are 3.57 times higher than in a human solution for an introductory question. The odds are reduced when we use the fine-tuned model to answer more complex questions. This means that a bug in a few-shot approach is half as likely as a bug in a human solution. 

\vspace{1mm}\noindent
\textit{\textbf{Finding 5:} The statistical tests demonstrate that LLM-generated solutions have fewer bugs than the solutions provided in the human dataset. Additionally, they contain fewer code smells than the solutions provided in the human dataset for most question types.}
\vspace{1mm}


\begin{figure}[!tp]
    \centering
\includegraphics[width=1\linewidth]{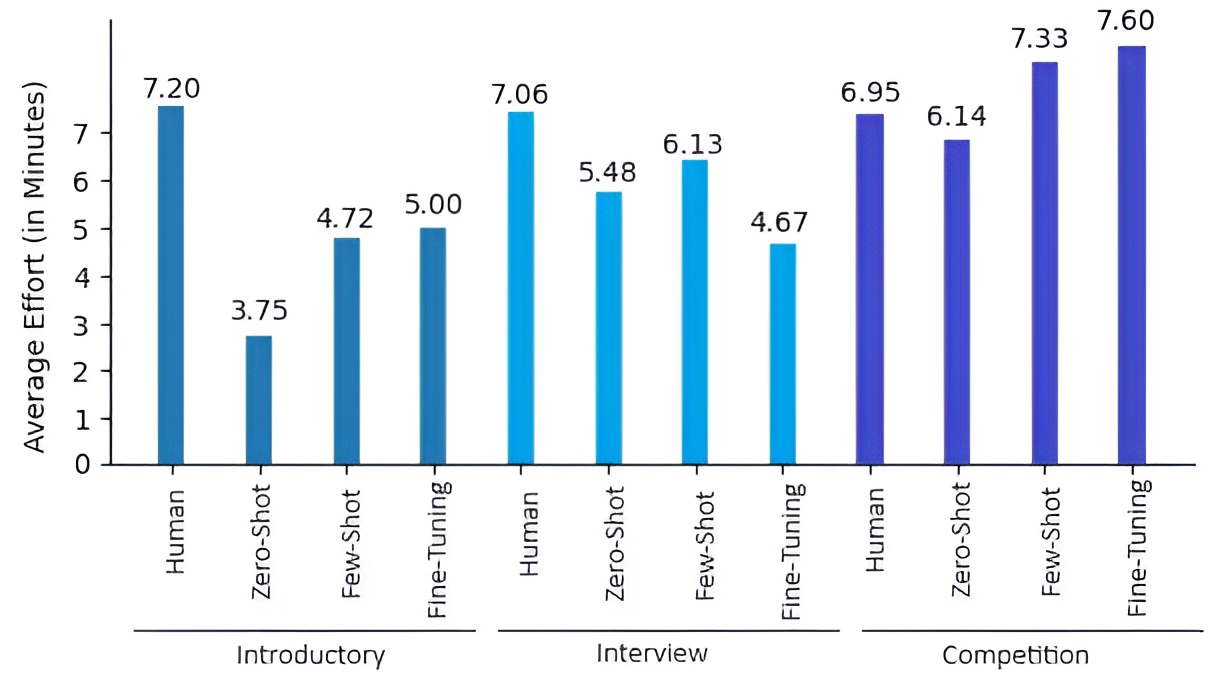}
    \caption{Average Effort in Minutes to Fix Bug or Code Smell.}
    \label{fig:effort_average}
\end{figure}

\subsection{Effort to Fix Issues}
SonarQube calculates the effort (in minutes) to fix the given issue. Figure \ref{fig:effort_average} shows the average time for each question type and prompt strategy. This figure demonstrates the average effort required to fix bugs or code smells across three different task categories for code generated by humans and by ChatGPT under the three prompting strategies. Each bar represents the average time one should spend (according to SonarQube) correcting the code in each condition. 
In the Introductory task category, human-written code required the most effort to fix (7.20 min), while LLM-generated code using zero-shot required the least (3.75 min), followed by few-shot (4.72 min) and fine-tuned (5.00 min). For Interview tasks, human code again requires the most effort (7.06 min), while fine-tuned (4.67 min) performed best among LLMs, followed by zero-shot (5.48 min) and few-shot (6.13 min). In Competition-level tasks, human code took 6.95 min to fix, less than few-shot (7.33 min) and fine-tuned (7.60 min), with zero-shot requiring the least time (6.14 min). Overall, LLMs reduced effort in simpler tasks, but this advantage diminished as the tasks became more complex.

While SonarQube’s effort estimates are approximate and may not capture all real-world debugging nuances, they serve as a valuable proxy for gauging the relative time and complexity involved in addressing code issues.

When comparing the effort in minutes using the Mann-Whitney U test, we found statistical differences between the three types of questions available and the human solutions with zero-shot and fine-tuned strategies, as shown in Table~\ref{tab:stat-efforttime}. Few-shot distribution was different for introductory and interview questions compared to human solutions. We observed a statistical difference between few-shot solutions and zero-shot solutions in effort time (Table \ref{tab:stat-efforttime} row FS x ZS). We rejected the \( H_{0} \) hypothesis of equal effort in minutes with \( p < 0.001 - introductory \), \( p < 0.001 - interview \), and \( p = 0.03 - competition \), respectively, observing Cliff's delta effect sizes as small, negligible, and negligible. When comparing the zero-shot and fine-tuned models, we did not observe a statistical difference. 

\vspace{1mm}\noindent
\textit{\textbf{Finding 6:} The results suggest that most solutions provided by the models demand less effort than the ones provided by the human dataset. }

\begin{table}[!tp]
\caption{Odds Ratio for Reliability (Bugs).}
\label{tab:stat-odds}
\begin{tabular}{llll}
\toprule
\multicolumn{1}{l|}{Test}                                 & introductory & interview & competition \\ 
\midrule
\multicolumn{1}{l|}{Zero-Shot x Human}                    & 0.008***  & 0.006***     & 0.0005***    \\
\multicolumn{1}{l|}{Few-Shot x Human }                     & 0.003***  & 0.006***     & 0.001***    \\
\multicolumn{1}{l|}{Fine-Tuning x Human }                  & 0.00***   & 0.019***     & 0.0005***   \\
\multicolumn{1}{l|}{Few-Shot x Zero-Shot }                 & 0.410      & 1.000         & 2.020        \\
\multicolumn{1}{l|}{Fine-Tuning x Zero-Shot}              & 0.000**     & 3.200       & 1.000       \\ 
\bottomrule
\multicolumn{2}{l}{p\textless{}0.05* p\textless{}0.01** p\textless{}0.001***} &              &           
\end{tabular}
\end{table}

\begin{table}[!tp]
\caption{Odds Ratio for Maintainability (Code Smells).}
\label{tab:stat-smells}
\begin{tabular}{llll}
\toprule
\multicolumn{1}{l|}{Test}                                 & introductory & interview & competition \\ 
\midrule
\multicolumn{1}{l|}{Zero-Shot x Human}                    & 3.57***  & 1.3     & 0.67    \\
\multicolumn{1}{l|}{Few-Shot x Human }                     & 1.29  & 0.80     & 0.52*    \\
\multicolumn{1}{l|}{Fine-Tuning x Human }                  & 0.00***   & 0.47     & 0.00***   \\
\multicolumn{1}{l|}{Few-Shot x Zero-Shot }                 & 0.36*      & 0.60         & 0.78        \\
\multicolumn{1}{l|}{Fine-Tuning x Zero-Shot}              & 0.00***     & 0.35**       & 0.00***       \\ 
\bottomrule
\multicolumn{2}{l}{p\textless{}0.05* p\textless{}0.01** p\textless{}0.001*** } &              &          
\end{tabular}
\end{table}

\begin{table}[!tp]
\caption{Mann-Whitney and Cliff's Delta for Effort Time.}
\label{tab:stat-efforttime}
\begin{tabular}{llll}
\toprule
\multicolumn{1}{l|}{Test}                                 &  introductory          &  interview  & competition       \\ \midrule
\multicolumn{1}{l|}{ZS x H}                    &  0.54 large*** & 0.21 small***       & 0.1 negligible*  \\
\multicolumn{1}{l|}{FS X H}                       & 0.38 medium*** & 0.14 negligible*** & 0.007 negligible  \\
\multicolumn{1}{l|}{FT x H}                        & 0.24 small & 0.27 small***  & -0.08 small     \\
\multicolumn{1}{l|}{FS x ZS}                  & -0.19 small*** & -0.08 negligible*** & -0.09 negligible* \\
\multicolumn{1}{l|}{FT x ZS}                   & -0.50 large  & 0.03 negligible & -0.22 small       \\ 
\bottomrule
\multicolumn{4}{l}{Human(H), Zero-Shot(ZS), Few-Shot(FS), Fine-Tuning(FT)} \\
\multicolumn{4}{l}{p\textless{}0.05* p\textless{}0.01** p\textless{}0.001***}
\end{tabular}
\end{table}


\subsection{Performance Analysis}

Table~\ref{tab:correctnesstable} presents the model performance analysis for 15 code solutions sampled from each prompt approach. These solutions were drawn from three difficulty levels: introductory, interview, and competition, with five code snippets per level. We executed the code snippets in our Python environment by running a \texttt{python3 file.py} command. We inserted inputs according to the test cases available in the problem description and checked the generated output. 



\begin{table}[!tp]
\centering
\caption{Pass@1 results.}
\label{tab:correctnesstable}
\begin{tabular}{l|c|c|c|c}
\toprule
Test & Human  & Zero-Shot  & Few-Shot & Fine-Tuned \\
\midrule
Pass@1 & 1.0 & 0.53 & 0.87 & 0.47 \\ 
\bottomrule
\end{tabular}
\end{table}



The results showed that the human-created solution outperformed the LLM-generated ones. 
Fine-tuned solutions, which have the best quality metrics, struggled to generate code for complex solutions. In general, the main issues observed were failure to pass all test cases and compilation errors. Going deeper, the solution generated for the problem 3827 was unable to handle the test case input and raised a runtime error during the execution: ``\textit{ValueError: invalid literal for int() with base 10}''. Solution 3839 insisted on returning the same (incorrect) output regardless of the test case input. One solution was unable to identify the end of the input and kept waiting for more data to arrive in problem 3878. We also observed cases where the ``return'' statement was out of place (786), functions were declared without a caller, repeating the input as the output (question 4353), syntax error (3827), incorrect results (3839, 3756, 3916, 260, 2568, 1871).
Few-shot solution issues were restricted to hardcoded inputs (instead of asking the user to input the test case data, in the interview question), and outputs did not match the expected result in a competition question. For zero-shot, the difference was that unexpected outputs were observed at all levels (introductory, interview, and competition), hard-coded inputs were present in the introductory and competition levels, and a syntax error was present in an introductory solution. 
Finally, some solutions returned partially correct outputs, passing only one or a few test cases. Partial solutions represented 47\% of the zero-shot solutions, and 13\% of the few-shot and fine-tuned solutions.

\vspace{1mm} \noindent
\textit{\textbf{Finding 7:} We found the odds of bugs (quality metric of reliability) are very low ($\leq 0.19\% $  ) compared with human solutions, while the odds of code smells are low mostly with more complex questions. We reject $H_{0}$ for solutions having the same effort time distribution for all comparisons between human and AI-generated solutions, except when comparing human x few-shot and human x fine-tuned for competition questions and human x fine-tuned for introductory questions. Few-shot performance outperformed the other AI methods in terms of Pass@1. } 
\vspace{1mm}

Finally, to aid reader understanding, we provide an example of an issue severity found by SonarQube in the dataset. In the human-written dataset, the following code solution was submitted as part of an interview-level task. 

\begin{lstlisting}[basicstyle=\small\ttfamily]
from itertools import *

k, p = list(map(int, input().split()))
ss = 0
for i in range(1, k + 1):
    s = str(i)
    num = int(s + ''.join(reversed(s)))
    ss += num
    ss %= p
print(ss)
\end{lstlisting}

SonarQube flagged a critical issue and generated the message: “Import only needed names or import the module and then use its members.” Specifically, the code includes an unnecessary import: it uses from itertools import * on line 1 but does not use any functions from the itertools module. This type of code smell highlights the inclusion of unused library imports. There are additional examples in our replication package \cite{SupplementaryMaterial}.

\section{Discussion} \label{discussion}

The analysis of code quality metrics and effort across different prompting strategies reveals several essential insights about the effectiveness and trade-offs of various approaches to AI-assisted code generation. Our findings have significant implications for software engineering. 

\textbf{Code Quality - LLMs vs. Humans:} The comparison between human-generated and AI-generated datasets revealed notable differences in code quality across multiple dimensions. The AI-generated datasets, particularly those created using fine-tuning strategies, consistently demonstrated improved metrics in terms of maintainability and reliability compared to the human-generated baseline. For example, fine-tuned models reduced the prevalence of high-severity issues, such as blocker and critical bugs, and shifted them toward lower-severity categories, like minor issues. This improvement highlights the ability of fine-tuned AI models to produce more robust and maintainable code solutions. On the other hand, human-generated datasets exhibited a higher proportion of critical and major issues, reflecting potential inconsistencies in coding standards and quality. However, in competition-level problems, the AI-generated solutions (few-shot) sometimes introduced unique structural critical issues not present in the human-generated code, emphasizing the need for critical evaluation and validation of AI outputs \cite{Florez2024}. These results underscore the potential for AI tools to complement human coding efforts, particularly in generating higher-quality code, which is similar to the results found by  Wadhwa et al. \cite{Wadhwa2024}, while revealing specific challenges that must be addressed to integrate these tools into educational and professional contexts fully.


\textbf{Maintainability vs. Reliability Trade-offs: }The distribution of issue types (maintainability vs. reliability) across different approaches reveals a critical pattern. While all AI-generated solutions showed improved reliability metrics compared to the baseline human dataset, the improvement in maintainability issues was less pronounced. This suggests that current AI models may better avoid functional bugs than address structural code quality concerns.

\textbf{Assessing Strategy Suitability for Code Generation:} Fine-tuned models demonstrated significant advantages regarding issue severity, particularly in reducing high-severity issues such as blockers and critical bugs. For example, fine-tuning consistently shifted severe issues to minor ones, as observed across all difficulty levels. This suggests that fine-tuned models can enhance the maintainability and reliability of code solutions, particularly in educational contexts where learning objectives align with real-world industry expectations. 

Another metric analyzed was the effort required to resolve the issues. AI-generated solutions, especially those created using fine-tuned models, needed less time to fix bugs for introductory and interview-level tasks, showcasing their ability to simplify coding processes in less complex scenarios. However, the effort required increased for competition-level tasks, possibly due to the inherent complexity of the problems and the nuanced structural issues introduced by the models. This highlights the need to carefully consider task complexity when choosing AI-assisted approaches \cite{bogomolov2022assessingprojectlevelfinetuningml4se,Negri-Ribalta2024}.

In contrast, the fine-tuned model failed to deliver the best performance, mainly for complex problems. One possible reason is that the number of questions used in the process was not sufficiently representative to efficiently update the model weights. Full fine-tuning can be extremely expensive to apply in contexts where the model size or the number of parameters is elevated \cite{weyssow2023exploring}. Limiting the number of samples can reduce costs, with the side effect of creating pitfalls in the model outcomes. This problem is more challenging with code generation datasets, as lexical similarity brings difficulties in sample selection \cite{tsai2024code}. Techniques to avoid costly full fine-tuning without penalizing the efficiency include parameter-efficient fine-tuning (PEFT) and data pruning \cite{tsai2024code}, which can be explored in future work. 

Manual analysis of failures indeed revealed that the problems are, in many cases, easy to fix, indicating that human intervention is still necessary; however, LLM support might be welcome. The performance of human-generated code was not a surprise. The APPS dataset is composed of solutions mined from websites for training purposes. Since the questions are not intended for use in a professional environment, the solutions may be prepared to be correct in the first place, leaving quality metrics as a low priority or even unobserved. 

\textbf{Impact when Learning How to Code: } \label{applications}LLMs can potentially transform coding education and practice by acting as supplementary tools. Educators can design assignments and projects where students critically analyze, debug, and improve AI-generated solutions, fostering practical problem-solving skills while reinforcing foundational programming concepts. Tools like SonarQube can be integrated into these activities, also suggested by Sofronas et al. \cite{knowledge3040036}, enabling students to assess their refinements and develop industry-relevant skills. For instance, fine-tuned models in this study consistently achieved better reliability metrics, offering a benchmark for teaching software quality and evaluation. Additionally, automatically categorized datasets can help create structured modules that simulate real-world challenges.

By integrating AI into curricula, educators can reduce their workload for generating coding questions and solutions, promote personalized learning, and provide students with real-time feedback. However, the study also emphasizes the need for students to validate AI-generated outputs, as LLMs can occasionally introduce structural issues. Teaching students to assess maintainability and reliability trade-offs critically prepares them to address these limitations effectively, while still adapting to the trend of using LLMs to generate code. Moreover, curricula should include AI-assisted software development, ethical considerations, and human-AI collaboration to prepare students for the evolving landscape of software engineering. Through these approaches, students can cultivate the judgment and skills necessary to balance the strengths of AI tools with their inherent limitations, fostering technical expertise and critical evaluation skills.

\subsection{Limitations of the Study} \label{limitations}

Cross-contamination may occur when instances have been exposed to specific coding patterns or prompts, which can influence subsequent responses from the LLM. This could affect the internal consistency of results when comparing zero-shot, few-shot, and fine-tuned prompts.
Prompt Sensitivity may be present due to the quality and accuracy of AI-generated code, which relies on the quality and accuracy of the provided prompts. Challenges in the wording of the prompt can lead to significantly different results, introducing variability in the experiment.

The baseline human dataset includes only Python solutions to specific problems, such as introductory, competition, and interview-level questions. This limits the generalizability of the findings to other programming languages or real-world software development scenarios. In addition, the study focuses on GPT-4 and a fine-tuned version, which may not represent the full spectrum of available LLMs. However, it is the most popular LLM. Another limitation of using the APPS data is that it lacks critical contextual information about the human-generated code. However, the APPS dataset is composed of questions gathered from relevant websites, including paid ones, and has been curated by a crowd of thousands of users. 

The sample analyzed for performance analysis may not accurately represent the population, potentially biasing the results. We randomized the selection and used a representative set of questions.  Additionally, we did not test how variations in LLM parameters or token limits affect performance. Instead, our focus was on comparing different prompting approaches and the difficulty of tasks. Future work could specifically explore the impact of parameter tuning.


Using SonarQube metrics, such as code smells and bugs, provides a standardized method for assessing code quality. Still, these metrics may not capture all aspects of software quality, and the effort estimation is limited to SonarQube's estimation algorithm, which can not reflect real-world developer effort.
Effort Estimation: SonarQube's estimation algorithm measures the effort required to fix issues. However, these estimates may not fully reflect real-world developer effort, as they do not account for contextual factors such as team experience or code familiarity. Due to resource constraints, the number of rows used for fine-tuning was limited. The fine-tuning results suggest that using more rows may be beneficial for model performance. 

\section{Conclusion and Future Work} \label{conclusion}
This study highlights the potential of LLMs in advancing software engineering. By analyzing the quality of human- and AI-generated code across three difficulty levels using SonarQube, our analysis reveals that compared to human-generated code, zero-shot, few-shot, and fine-tuning approaches reduce high-severity issues, such as bugs and code smells, in the code. AI-generated code may still introduce less critical issues, such as minor code smells or stylistic inconsistencies, which do not require immediate attention and do not impact the functionality or correctness of the code. Our results showed that AI-generated solutions, remarkably fine-tuned models, shifted severe issues (e.g., blocker or critical) to less critical categories, such as minor issues. This improved overall code quality while still leaving room for minor refinements. However, the low performance of the fine-tuned model solutions sheds light on the limitations of this method, which is more expensive than zero- and few-shot prompting. 

Future work may explore the trade-off between quality, performance, and costs in different domains as well as advanced fine-tuning strategies like retrieval-augmented generation (RAG) and use different LLMs in the experiment while seeking diverse performance metrics. Expanding analyses to multiple languages and new quality metrics. Adding a component to the architecture to check the performance of the question is another interesting future work that could advance this research toward a tool for automatically building a database of questions. Additionally, studying the relationship between AI-generated code quality and educational outcomes, ethical implications, and scalable frameworks for real-time feedback could also support the integration of LLMs into industry practice and software engineering education.


\balance
\bibliographystyle{IEEEtran} 
\bibliography{main.bib}
\end{document}